\documentclass[aps,reprint,groupedaddress,bibliography]{revtex4-1}

\usepackage{siunitx}
\usepackage{empheq}
\usepackage{color}
\usepackage{footnotehyper} 
\usepackage{balance}
\usepackage{graphicx} 
\usepackage[english]{babel}
\usepackage{amssymb}
\usepackage{hyperref}

\definecolor{cream}{RGB}{222,217,201}
\definecolor{green2}{rgb}{0.00,0.27,0.13}

\begin{document}

\title{Collective diffusion coefficient of a charged colloidal dispersion: \\ interferometric measurements in a drying drop}

\author{Benjamin Sobac$^a$}
\email{bsobac@ulb.ac.be}
\author{Sam Dehaeck$^a$}
\author{Anne Bouchaudy$^b$}
\author{Jean-Baptiste Salmon$^b$}
\email{jean-baptiste.salmon-exterieur@solvay.com}

\affiliation{$^a$ TIPs lab, Universit\'e libre de Bruxelles, 1050 Brussels, Belgique\\
$^b$ CNRS, Solvay, LOF, UMR 5258, Univ. Bordeaux, F-33600 Pessac, France}



\begin{abstract}
In the present work, we use Mach-Zehnder interferometry to thoroughly investigate the drying dynamics of a 2D confined drop of a charged colloidal dispersion. This  technique makes it possible to measure the colloid concentration field during the drying of the drop at a high accuracy (about $0.5$\%) and with  a high  temporal and spatial resolution  (about $1$ frame/s and $5$~$\mu$m/pixel). These  features allow us to probe mass transport of the charged dispersion in this out-of-equilibrium situation. In particular, our experiments provide the  evidence that mass transport within the drop can be described by a purely diffusive process for some range of parameters for which the buoyancy-driven convection is negligible. We are then able to extract from these experiments the collective diffusion coefficient  of the dispersion $D(\varphi)$ over a wide concentration range $\varphi=0.24$--$0.5$, i.e.\ from the liquid dispersed state to the solid glass regime, with a high accuracy. The measured values of $D(\varphi)\simeq 5$--$12 D_0$ are significantly larger than the simple estimate $D_0$ given by the Stokes-Einstein relation, thus highlighting the important role played by the colloidal interactions in such dispersions. 
\end{abstract}

\maketitle


\section{Introduction}
Understanding mass transport phenomena in complex fluids, and in particular in colloidal dispersions, does not only raise many fundamental issues, but is also relevant for optimizing many processes, such as the drying step of liquid films in the coating industry~\cite{Keddie}, formation of microparticle powders using spray drying in the food and pharmaceutic industries~\cite{Lintingre2016}, or membrane ultra-filtration in waste water treatment~\cite{Shannon:08}. In the drying process of a colloidal dispersion for instance, diffusion opposes to the formation of concentration gradients induced  by drying, and the concentration dynamics of the colloids depends quantitatively on the collective diffusion coefficient $D$ of the dispersion~\cite{Routh:13}.  This transport coefficient relates through the Fick's law, the colloid density gradient  to the diffusive flux of colloids due to their coordinated motion. This coefficient therefore differs from the self-diffusion coefficient $D_{\rm{self}}$ related to the Brownian movement of a tagged colloid in the dispersion~\cite{Nagele,Russel}.
In most cases, $D$ is approximated by $D_0$,
 the value given by the Stokes-Einstein relation,  only valid to describe both self-diffusion and collective diffusion in very dilute dispersions as it does not include  both colloidal and hydrodynamic interactions.  
These interactions are particularly important for charged-stabilized dispersions, especially at low salinity and high colloid concentration, possibly leading to values of $D$ significantly higher than $D_0$  (and $D_{\rm{self}}$) because of the highly coordinated motion of the colloids due to the 
long-range repulsive interactions~\cite{Nagele}. Measurements of the collective diffusion coefficient of charged dispersions were obtained using dynamic scattering techniques 
(light and x-ray), and were successfully compared  to  advanced models~\cite{Gapinski:07,Bowen:00,Petsev92}, but only for highly monodisperse systems in a rather dilute regime. Indirect estimates were also made using permeate flux measurements in ultrafiltration experiments~\cite{ROA2016,Bowen2007}, again in a dilute regime. However,  measuring such a coefficient at high colloid concentration remains an experimental challenge, mainly because charged dispersions often form soft solids at volume fractions below the colloid close packing due to  the long-range repulsions. The theoretical description of collective diffusion in this concentrated regime is still an open issue, and could also include poroelastic contributions, as recently suggested in the context of drying~\cite{Style:11}. Very few experimental studies have reported direct measurements of $D$ in such regimes~\cite{Loussert:16,Goehring2017},  while these measurements are 
still necessary to understand the phenomena at stake and test models.

\begin{figure}[htbp]
\begin{center}
\includegraphics[width=8cm]{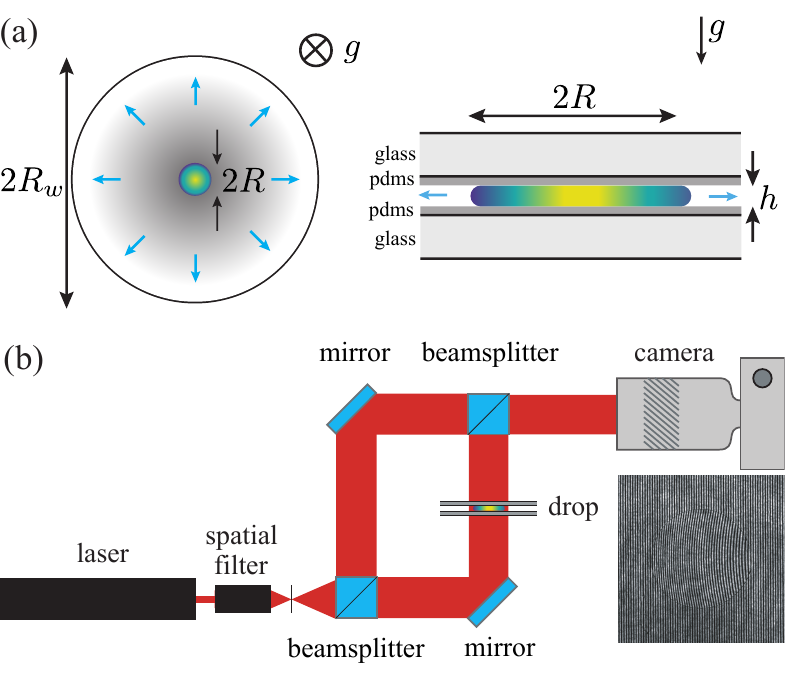}
\caption{Schematic illustration of the experimental setup involving (a)  a confined drying cell in which a 2D drop of a colloidal dispersion evaporates and (b) a Mach-Zehnder interferometer. The latter enables to measure, from patterns  of interference fringes (referred to as interferograms), the particle concentration field in the drop. A typical interferogram is shown at the bottom right corner. Our experiments involve Ludox~AS 
as colloidal dispersion, drops of initial radius $R_0= 1$--$1.7$~mm, and a confined cell with $R_w = 38.1~$mm and $h=150$ or $300~\mu$m.\label{fig:Setup}}
\end{center}
\end{figure}

Daubersies {\it et al.}\ have developed an original method to measure the collective diffusion coefficient of an aqueous binary mixture (water $+$ non-volatile solute), even in concentrated regimes~\cite{Daubersies:12}. This method is based on the drying of a drop of complex fluid in a confined geometry. Their experiment is schematically illustrated in Fig.~\ref{fig:Setup}a: a drop of a few microliters of the fluid is observed drying between two circular, transparent wafers covered with 
an appropriate coating to prevent the contact line from sticking.
In this confined geometry (typically $h\approx100-300~\mu$m), the drying time of a water drop of initial radius $R_0 \simeq 1$~mm is a few hours at ambient conditions for $R_w=38.1$~mm~\cite{Clement:04}.  Such an  experiment enables an easy observation while providing controlled drying conditions, where complex phenomena related to the free surfaces are limited. Indeed, the axisymmetrical drying of the drop, combined with the 
 small free surface ($\simeq 2 \pi R h$) prevents flows such as   capillary-induced convection or Marangoni-induced convection. 
For these reasons, many groups have carried out such experiments, in particular to explore the phase diagram of polymer solutions~\cite{Daubersies:12}, to study the drying of hard-sphere~\cite{Leng:10} and charge-stabilized dispersions~\cite{Loussert:16,Bouchaudy:2019}, or to investigate the mechanical instabilities associated with the formation of solid crusts~\cite{Boulogne:13,Yunker:12}. 

In this geometry (see~Fig.~\ref{fig:Setup}a), the evaporation of the solvent leads to the formation of concentration gradients within the drop due to the competition between the drop shrinkage and solute diffusion. Then, by measuring both the drying rate and the solute concentration along the drop diameter using Raman confocal spectroscopy, Daubersies {\it et al.}\ estimated the collective diffusion coefficient $D$, in their case, of an aqueous co-polymer solution~\cite{Daubersies:12}. Later, Loussert {\it et al.}\ applied the same method to a commercial dispersion, namely Ludox AS
, made of charged silica nanoparticles dispersed in water~\cite{Loussert:16}. This dispersion is known to exhibit a transition  from a liquid dispersed state to a soft repulsive solid, at a colloid volume  fraction $\varphi \simeq 0.32$, as shown by Bouchaudy {\it et al.},  who performed confined drying experiments with an original setup making possible the measurements of mechanical drying-induced stresses~\cite{Bouchaudy:2019}. The experiments of Loussert {\it et al.}\ provided the first direct measurements of $D$ for such charge-stabilized dispersion over a wide range of concentration, $\varphi\simeq 0.24$--$0.55$. Similar measurements were also performed by Goehring {\it et al.}\ on similar dispersions (Ludox SM, HS and TM) using spatially-resolved small-angle X-ray scattering (SAXS) 
 but in a uni-directional drying cell~\cite{Boulogne:14,Goehring2017}.  Both experiments  showed that the collective diffusion coefficient of such dispersions is significantly larger than $D_0$ by a factor ranging from $\simeq 10$ to $40$, highlighting the important role played by colloidal interactions. Moreover, both experiments showed striking dependencies of $D$ with the colloid concentration $\varphi$ that are not explained by the current models~\cite{Loussert:16,Goehring2017}. 

The experimental techniques used in the above-mentioned methods (Raman confocal spectroscopy~\cite{Loussert:16} and SAXS~\cite{Goehring2017}, see also refs.~\cite{Merlin:11,Boulogne:14,Yang2018}) do not allow easy measurements of the concentration fields with  a  simultaneously  high temporal and spatial resolution. Higher resolutions are nevertheless necessary to get more accurate values of $D$, and thus to go further in our understanding of the phenomena at play. In the present work, we combined the method of the drying of a confined drop with a visualization by Mach-Zehnder interferometry, as illustrated in Fig.~\ref{fig:Setup}b. This optical technique enables to measure the colloid concentration field over the whole drop with a high absolute accuracy, typically $\pm 0.5\%$, and with a high temporal and spatial resolution, typically 1 frame/s and $\simeq 5~\mu$m/pixel. These unique features allow accurate measurements of the collective diffusion coefficient, here provided for the aqueous colloidal dispersion Ludox AS
, for direct comparison with the measurements of ref.~\cite{Loussert:16}. 

The present paper is organized as follows. In Section~\ref{sec:ConfinedDryingTheory}, we first briefly recall the theoretical description of the drying of a dispersion in this confined geometry. In \textit{Materials and Methods} (Section~\ref{sec:MM}), we present the dispersion under study, including rheological measurements evidencing a transition to a soft repulsive solid at a  volume fraction $\varphi \simeq 0.33$, and detail both the 
interferometric setup and the method we used to extract concentration fields. Then, Section~\ref{sec:Results} reports precise measurements of $D(\varphi)$ over a wide range of concentration $\varphi \simeq 0.24$--$0.5$. Numerical solutions of the solute mass balance equation based only on diffusion correctly describe our measurements, hence demonstrating the robustness of our  data. In addition, measurements performed in a cell with a larger height ($h=300~\mu$m vs.\ $h=150~\mu$m), allow us to reveal the importance of buoyancy-driven free convection in such experiments. These results are rationalized 
by a model that enables to delimit the range of parameters for which measurements of $D$ are possible knowing density gradients and the viscosity of the dispersion. Finally, we discuss the implications of our work in Section~\ref{sec:discconc}.

\section{Confined drying of a colloidal dispersion \label{sec:ConfinedDryingTheory}}

The typical experiment of confined drying we consider is displayed in Fig.~\ref{fig:Setup}a: a drop of a few microliters is squeezed between two circular wafers of radius $R_w$, and separated by a  fixed height $h$. This drop dries in the air under ambient conditions and shrinks axisymmetrically, the contact line receding freely due to the presence of specific coatings on the wafers. The theoretical description of the drying dynamics of a binary mixture (solvent $+$ non-volatile solute) in this geometry has been fully described in ref.~\cite{Daubersies:11} (see also ref.~\cite{Loussert:16,Bouchaudy:2019} for the specific case of colloidal dispersions).
The water evaporation from an aqueous dispersion being quasi-steady and limited by the diffusion of the vapor towards the edge of the cell, the radius of the drop varies as~\cite{Daubersies:11}:
\begin{eqnarray}
\frac{\text{d}R }{\text{d}t} = \frac{D_s V_s c_{\rm{sat}}(1-\text{R.H.})}{R \ln(R/R_w)}\,, \label{eq:dryindynamics}
\end{eqnarray}
with $D_s$ the diffusion coefficient of the water vapor in air, $V_s$ the molar volume of liquid water, $c_{\rm{sat}}$ the concentration of the saturated vapor pressure, and $\text{R.H.}$ the external relative humidity.  This equation remains valid for colloidal dispersions as long as the drop remains cylindrical with $R(t) > h$. Note that such a description implies an isothermia of the system at the ambient temperature $T$. Using the initial condition $R(t=0)=R_0$, the  analytical   solution of eqn~(\ref{eq:dryindynamics}) writes~\cite{Clement:04}:
\begin{eqnarray}
\frac{t}{\tau_f} = 1 -\frac{\alpha \left(\ln(\beta\alpha)-1\right)}{\ln(\beta)-1}\,, \label{eq:soldryindynamics}
\end{eqnarray}
with $\alpha = (R(t)/R_0)^2$, $\beta = (R_0/R_w)^2$ and
\begin{eqnarray}
\tau_f = \frac{R_0^2}{4 D_s V_s c_{\rm{sat}}(1-\text{R.H.})}(1-\ln(\beta))\,,  \label{eq:dryingtimes}
\end{eqnarray}
the drying time of a pure water drop of the same initial radius, evaporating in the same conditions.

The colloidal concentration field inside the drop is considered governed by a competition between the shrinkage of the drop induced by evaporation and diffusive mass transport. For simplicity, we will assume that the charge-stabilized dispersion can be described by a binary mixture only (i.e.\ colloids of radius $a$ dispersed in water), skipping the dynamics of molecular ionic species in the dispersion (namely salts and counter-ions), see later for discussions.
Denoting ${\bm v_c}$ and ${\bm v_s}$ the average velocities of the (incompressible) colloids and solvent,
conservation equations for these two quantities are:
\begin{eqnarray}
&&\frac{\partial \varphi}{\partial t }+ \bm \nabla . (\varphi {\bm v_c})=	 0\,, \\
&&\frac{\partial (1-\varphi)}{\partial t }+ \bm \nabla . (1-\varphi) {\bm v_s}=	 0\,. 
\end{eqnarray}
If one defines the volume-averaged velocity of the mixture as ${\bm v} = \varphi \bm v_c + (1-\varphi) {\bm v_s}$, and the collective diffusion $D(\varphi)$ of the mixture as:
\begin{eqnarray}
\varphi\bm v_c = \varphi\bm v - D(\varphi) \bm \nabla \varphi\,,
\end{eqnarray}
then the 
global and colloid mass balances read in this axisymmetrical geometry~\cite{Loussert:16,Bouchaudy:2019}:
\begin{eqnarray}
&&\bm \nabla . \bm v = 0\,, \label{eq:divv}\\
&&\frac{\partial \varphi}{\partial t }+ \bm v .\bm \nabla \varphi = \frac{1}{r}\frac{\partial}{\partial r} \left( r D(\varphi) \frac{\partial \varphi}{\partial r}\right)+\frac{\partial}{\partial z} \left(D(\varphi) \frac{\partial \varphi}{\partial z}\right)\,. \label{eq:TransportConv}
\end{eqnarray}

Water evaporation induces the shrinkage of the drop, leading thus to radial concentration gradients.
These gradients could in turn generate buoyancy-driven free convection (whatever the height of the cell)  as the colloids and water do not have generally the same density.
Free convection has been observed several times in such experiments, for molecular solutions (e.g.\ salts, glycerol, polymers~\cite{Selva:12,Lee:14,Pradhan2018,Daubersies:12}), and even colloidal dispersions at low volume fractions~\cite{Loussert:16,Bouchaudy:2019}. However, if we assume, as in ref.~\cite{Loussert:16}, that free convection does not significantly impact diffusive mass transport, then concentration profiles are almost homogeneous over $h$,  i.e.\ $\partial_z \varphi \simeq 0$.  Averaging   the global mass balance  eqn~(\ref{eq:divv}) over the height $h$ yields  $\int_0^h v_r(z,t) \text{d}z=0$ with $v_r$ the radial velocity of the mixture, and 
the same averaging of the colloid mass balance eqn~(\ref{eq:TransportConv}) leads  finally  to the diffusion equation:
\begin{eqnarray}
&&\frac{\partial \varphi}{\partial t }= \frac{1}{r}\frac{\partial}{\partial r} \left( r D(\varphi) \frac{\partial \varphi}{\partial r}\right)\,, \label{eq:Transport}
\end{eqnarray}
with the boundary condition  at the receding free surface: 
\begin{equation}
- \left.D(\varphi) \frac{\partial \varphi}{\partial r} \right|_{r=R(t),\,t} = \varphi(R(t),t)\frac{\text{d}R }{\text{d}t}\,, \label{BC1} 
\end{equation}
that ensures the non-volatility of the colloids. 
In the present work, we will return in more detail to the hypothesis of negligible free convection  that allow us to move from  eqn~(\ref{eq:TransportConv}) to eqn~(\ref{eq:Transport}),  see later Section~\ref{sec:roleofbuoyancy}  for details.

It also follows from the colloid mass balance  that  the average concentration in the drop $<\varphi>$ is related to the normalized area $\alpha = (R(t)/R_0)^2$ through:
\begin{eqnarray}
<\varphi> = \frac{1}{\pi R(t)^2}\int_0^{R(t)} 2\pi r \varphi(r,t) \text{d}r = \frac{\varphi_0}{\alpha}\,, \label{eq:phimoy}
\end{eqnarray}
where $\varphi_0=\varphi(r,t=0)$ is the initial concentration in the drop.

The above eqn~(\ref{eq:Transport}) and (\ref{BC1}) 
can be written in a dimensionless form using the dimensionless variables $\tilde{t} = t / \tau_f $ and $\tilde{r} =r/R_0 $, and defining $D(\varphi) = D_0 \hat{D}(\varphi)$ with $D_0$ given by the Stokes-Einstein relation 
\begin{equation}
D_0 = \frac{k_B T} {6\pi \eta_s a}\,, \label{eq:D0}
\end{equation}
\noindent where $k_B$ is the Boltzmann's constant and $\eta_s$ is the water viscosity. 
This  set of two dimensionless equations only depends on a unique parameter (see Appendix A):
\begin{eqnarray}
\text{Pe} = \frac{R_0^2}{D_0 \tau_f} = \frac{4 D_s V_s c_{\rm{sat}}(1-\text{R.H.})}{D_0}\frac{1}{1-\ln(\beta)}\,. \label{eq:Pe}
\end{eqnarray}
This  P\'eclet number compares the characteristic time scale of diffusion of the colloids across the drop, $R_0^2/D_0$, with the drying time 
$\tau_f$. Thus, $\text{Pe} \ll 1$ corresponds to a drying with an almost homogeneous colloid concentration field, whereas for $\text{Pe} \gg 1$, sharp concentration gradients develop at the receding meniscus~\cite{Daubersies:11}.
Note that, due to the 2D configuration of the drying cell, $\text{Pe}$ weakly depends on the wafer and drop radii, see the logarithmic term in eqn~(\ref{eq:Pe}). This differs from the P\'eclet number derived in the 3D configuration of a drying spherical drop that is independent of any length scales~\cite{Sobac2019}.
The full dimensionless model (provided in Annex~A with its numerical resolution) permits to reveal all the relevant variables of the confined drying experiment, namely the normalized area $\beta$ (and thus the initial radius $R_0$ for a fixed $R_w$), 
the drying time $\tau_f$, the P\'eclet number $\text{Pe}$, and the initial concentration of the drop $\varphi_0$.

\section{Materials and methods \label{sec:MM}}

\subsection{Colloidal dispersion}

We used the commercial aqueous dispersion of silica nanoparticles referred to as Ludox AS40 
(monodisperse anionic grade, pH=9.1, Sigma Aldrich). The mass fraction of the stock dispersion is about 40\% and the mean radius of the colloids is about $a\simeq11~$nm. The nanoparticles are negatively charged because of ionized silanol groups at their surface, with ammonium hydroxide as counterions. 
The bare surface charge of these silica particles depends on the pH of the dispersion (as also revealed by electrophoretic mobility measurements in the dilute regime, see e.g.\ ref.~\cite{Ning:08}), and typical values of $0.5$--$0.6$~e/nm$^2$ are commonly reported using titration experiments at pH$=9$~\cite{Jonsson:11}. 
The volume fraction $\varphi_0$ of the stock dispersion was estimated by using measurements of its dry extract ($120^{\circ}$C for 30~min), and of its density, assuming an ideal mixture. We found $\rho_c = 2.216$~g/mL for the density of the silica particles and $\varphi_0\simeq0.24$. We performed experiments involving this Ludox AS dispersion with several initial volume fractions $\varphi_0\simeq0.24$, $0.14$ and $0.05$. These two latter volume fractions were simply obtained by dilution of the stock dispersion with deionised water. 

At $T=21^\circ$C, the Stokes-Einstein relation eqn~(\ref{eq:D0}) predicts a diffusion coefficient $D_0 \simeq 2\times 10^{-11}~$m$^2$/s for the Ludox AS. As collective diffusion depends \textit{a priori} on colloidal interactions, we expect that our measurements of the collective diffusion coefficient $D(\varphi)$ strongly differ from $D_0$, at least in concentrated regimes. It should also be noted that the ionic content of the commercial batch is not precisely known. 
However, preliminary experiments carried out with dialysed dispersions, as well as comparisons of the observed behaviours with some reported works~\cite{Boulogne:14,Goehring2017} suggest that the commercial dispersion contains only a small amount of ionic species in solution, probably of the order of a few mM or less.  
Hence, great care should be taken for interpreting our results with theoretical models of $D(\varphi)$, in particular when the electrostatic contribution 
for the inter-particles interactions is considered,  see Section~\ref{sec:discconc}. 
 The choice of such a dispersion for our study is mainly motivated by performing a direct comparison of our measurements of $D(\varphi)$ with those reported in ref.~\cite{Loussert:16} (see Section~\ref{sec:Loussert}). 

\begin{figure}[ht]
\begin{center}
\includegraphics[width=8cm]{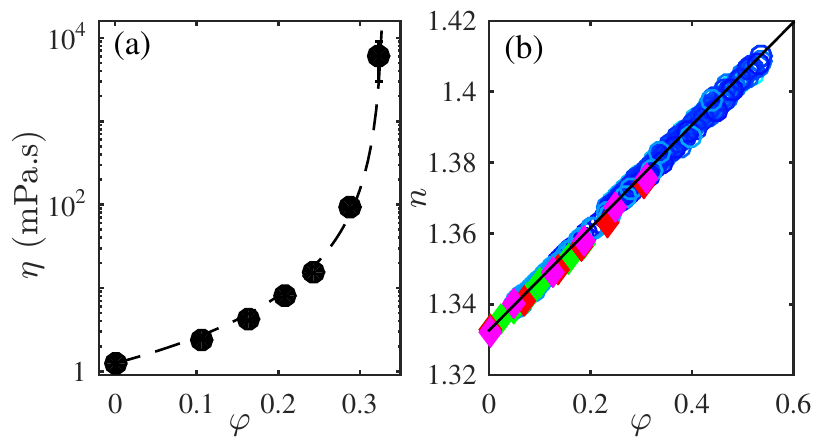}
\caption{(a) Viscosity of the dispersion $\eta$ as a function of the colloid volume fraction $\varphi$. The experimental points ($\bullet$, see Annex~B) are fitted by the rheological law $\eta = \eta_s(1-\varphi/\varphi_c)^{-2}$ with $\varphi_c = 0.33$ in dashed line. (b) Refractive index of the  dispersion  $n$ vs.\ $\varphi$. The experimental data, obtained either classically using a refractometer for Ludox AS ($\color{red}{\blacklozenge}$), Ludox TM  ($\color{magenta}{\blacklozenge}$), Ludox TMA~\cite{Ning:08} ($\color{green}{\blacklozenge}$), or from our confined drying experiments of Ludox AS using a Mach-Zehnder interferometer (${\circ}$ with a gradient of blue colors for $10$ experiments, see Section~\ref{sec:protocolinterf}), are all superimposed and fitted by the linear law $n = 0.145\varphi + 1.3362$ in solid line.
\label{fig:FlowCurve}}
\end{center}
\end{figure}

The mass transport in a colloidal dispersion being possibly linked to its rheology, we characterized the rheological behavior of the dispersion by measuring its shear viscosity $\eta$ as a function of the colloid concentration $\varphi$. The result is displayed in Fig.~\ref{fig:FlowCurve}a, while methodological details for obtaining these data are provided in Annex~B. The rheological curve shows that the viscosity of the dispersion increases with the colloid concentration until it diverges at $\varphi_c \simeq 0.33$. Our rheological characterization strongly suggests that the dispersion evolves from a viscous Newtonian fluid towards a solid glass with a transition at this critical concentration $\varphi_c$, hence well-before the close-packing fraction. Note that the experimental points reported in Fig.~\ref{fig:FlowCurve}a appear rather well-fitted by the standard rheological law $\eta = \eta_s(1-\varphi/\varphi_c)^{-2}$ classically employed for hard-sphere suspensions~\cite{Guazzelli2018}. All these observations are in line with those reported by \citet{DIGIUSEPPE2012} on a similar system (Ludox HS40) evidencing also a transition to a solid glass at $\varphi \simeq 0.32$, as well as with other indirect observations indicating the appearance of a yield stress for such dispersions at similar volume fractions~\cite{Loussert:16,Goehring2017,Bouchaudy:2019,Boulogne:14}.

\subsection{Confined drying experiments}

We use glass wafers ($3$ inches in diameter, $1$~mm thick) spin-coated by a thin layer of cross-linked poly(dimethylsiloxane) PDMS (thickness $\simeq 30~\mu$m, Sylgard 184). This hydrophobic coating prevents the pinning of the receding meniscus during the drying of the drop. Spacers consist in small pieces (of a few mm$^2$) made of a glass slide of thickness $150~\mu$m, which are positioned close to the edge of the cell. Their thicknesses were measured using a 3D laser scanning confocal microscope (Keyence VK-X200 series). In the present study, the spacing of the drying cell is either $h=150~\mu$m or $h=300~\mu$m, the latter being simply obtained by the superposition of two spacers.

In a typical experiment, a drop of colloidal dispersion of a volume $V\simeq 0.5$--$1.5~\mu$L is pipetted at the center of one of the glass wafers and the cell is carefully closed using the second wafer. If necessary, the whole cell can be displaced using a manual stage to center the drop within the field of view of the acquisition setup. The drop  dries in ambient air whose temperature $T$ and relative humidity $\text{R.H.}$ remain roughly constant during an experiment. In the present work, we performed and analyzed 
experiments carried out in two different ambient conditions: [$\text{R.H.} \simeq 0.5$ and $T \simeq 21^\circ$C] or [$\text{R.H.} \simeq 0.23$ and $T \simeq 22.5^\circ$C]. Based on the definition of the P\'eclet number in eqn~(\ref{eq:Pe}), it corresponds for our dispersion to $\text{Pe} \simeq 5.3$--$6.2$ and $\text{Pe} \simeq 10.7$--$11.6$, respectively. These values of $\text{Pe}$ being larger than 1 suggests the presence of significant concentration gradients during drying (if the Stokes-Einstein value $D_0$ correctly describes the diffusive transport).

\begin{figure*}[htbp!]
\begin{center}
\includegraphics[width=16cm]{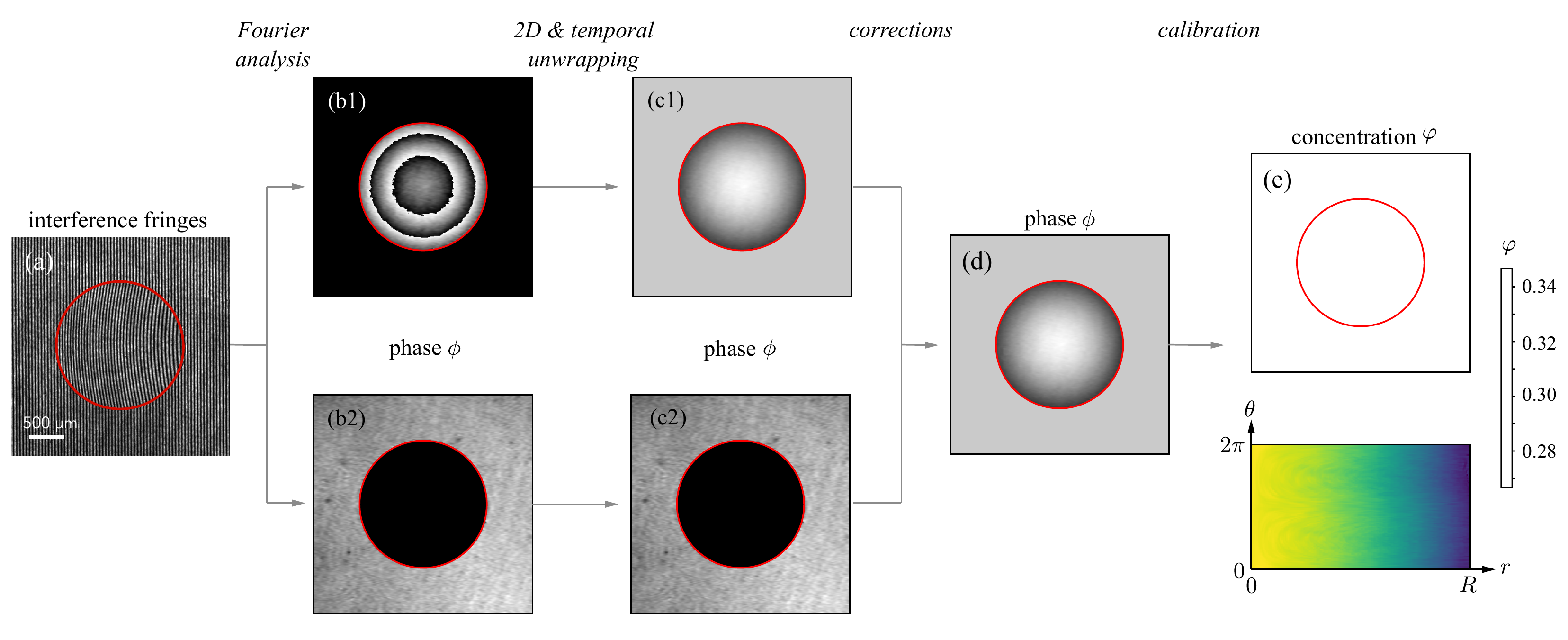}
\caption{Schematic view of all steps of the image processing enabling to measure the colloid concentration field within a drying confined drop from the finite-fringe interferograms. These various steps are presented in detail in Section \ref{sec:protocolinterf}. See also the Supplementary Movies 1 and 2, related to (a) and (e), respectively. \label{fig:Method}}
\end{center}
\end{figure*}

\subsection{Mach-Zehnder interferometer setup \label{sec:MachZehnder}}

The confined cell is placed in a Mach--Zehnder interferometer as shown in Fig.~\ref{fig:Setup}b. This optical technique enables us to visualize and measure the 2D colloid concentration field within the drying drop, making use of the concentration dependence of its refractive index $n$ (see Fig.~\ref{fig:FlowCurve}b and Section~\ref{sec:protocolinterf}). A He--Ne laser beam (wavelength $\lambda=632.8~$nm) is expanded and cleaned using a spatial filter, towards a first beamsplitter, which splits it into a reference beam and a measurement beam. The reference beam passes through the ambient air while the measurement beam passes through the cell. Finally, the two beams are recombined by a second beamsplitter and imaged on the CMOS camera (IDS UI-3040CP-v2, $1448\times1086$~pixels) by a Micro-Nikkor 105~mm f/2.8G objective. Due to the refractive index differences in the drop, the optical paths of the two beams are different and an interference pattern is recorded. A typical interferogram is shown in the bottom right corner of Fig.~\ref{fig:Setup}b. Note that the tiny circular black strip observed in the interferogram corresponds to the air--dispersion interface, the meniscus shape deflecting the measured beam out of reach of the objective. Moreover, for image processing purposes, the interferometer is set to generate a homogeneous system of vertical fringes before drop deposition, as still observed outside of the drying drop in the interferogram. The typical acquisition frequency in our experiments is $1$ frame per second, and the spatial resolution of our images is $\simeq 5~\mu$m per pixel. Note that while the image acquisition is started prior the deposition of the drop, the image processing begins once the axisymmetrical drop is within the camera field of view ($\simeq 5.4 \times 5.7$~mm$^2$). The typical elapsed time between the drop deposit and the start of the image processing is of the order of $15$ to $30~$s.

\subsection{Image processing \label{sec:protocolinterf}}

First, we perform a simple image analysis to extract the position and the size of the drop during drying. We proceed by fitting the thin circular black strip corresponding to the meniscus by a circle to obtain for each image the location of the drop center and the drop radius $R(t)$. From this, the drop area $A(t)=\pi R(t)^2$  and therefore the average colloid concentration in the drop $< \varphi >=\varphi_0 A_0/A(t)$, see eqn~(\ref{eq:phimoy}), can easily be deduced, as the initial drop radius $R_0$ and volume $V\simeq h \pi R_0^2$. 

Second, we analyse these finite-fringe interferograms based on the classic Fourier-transform algorithms from literature~\cite{takeda1982,kreisInterferomFFT}  (see Fig.~\ref{fig:Method} to view all of the processing steps). However, in contrast to a classic approach where a reference interferogram (taken before the start of the process) is subtracted in order to account for optical distortions inherent to the physical setup~\cite{kreisInterferomFFT,dehaeckEvapCocktails}, such a reference is not easily obtained here. Indeed, it was found that the deposition of the drop followed by the closing of the cell already lead to a significant distortion of the refractive index field near the drop edge (see Fig.~\ref{fig:Sequence_AS40normal4}a). As such, no satisfactory reference image could be obtained. Nevertheless, due to the high quality optics used throughout the setup, the obtained phase images were nearly homogeneous in the limited field of view that was chosen, even without such a reference subtraction (see Fig.~\ref{fig:Method}b2).

Obtaining the phase image $\phi$ is  only the beginning however, the next step is performing the  2D and temporal phase unwrapping (as a phase image is limited to values of $-\pi$ to $\pi$). Due to the presence of two distinct refractive index regions (dispersion and air) and a moving interface across which the interference signal is lost, the unwrapping algorithm needs to be applied separately to both domains. This is achieved by applying two complementary masks to the phase image, once on the outside of the drop (see Fig~\ref{fig:Method}b1) and once on the inside of the drop (see Fig~\ref{fig:Method}b2). On these two masked images, the unwrapping algorithm of \citet{HerraezUnwrapping} is used. While this will eliminate any phase jumps inside a single image (see Fig.~\ref{fig:Method}c1 and \ref{fig:Method}c2), there could still be phase jumps over time. To get some noise resistance, a suitably chosen small group of pixels was averaged over to detect such phase jumps, and any detected phase jump was then applied to the whole (masked) unwrapped phase image. Note that these masks were constructed for each image based on the detection of the meniscus position mentioned at the beginning of this section.

At this point, a fully-processed phase image is obtained for both domains, which is already accurate in a relative sense. However, in order to improve the absolute accuracy over longer time periods, one needs to take care of several issues. Random phase jumps from one image to the next, due to table vibrations or other influences, are quite natural as our interferometer does not use a common path for both beams. While these jumps average out over time, there is also typically a drift of the phase visible when measuring over longer time periods, presumably due to drifts in the laser beam pointing or wavelength stability. Fortunately, these two phenomena affect both regions equally and uniformly (over a sufficiently small field of view). Now, as the anticipated refractive index changes in the air should be negligibly small, one can extract a phase correction for each image (with respect to the first image in the set) so as to keep the phase constant in the air region. This correction is then also applied to the drop region (see Fig.~\ref{fig:Method}d). 

The next issue to solve is the transformation from phase into refractive index and subsequently into concentration. The first step follows from $\Delta n=\lambda\Delta\phi/(2\pi h)$, valid for Mach-Zehnder interferometry with a Hele-Shaw cell~\cite{dehaeckEvapCocktails}. 
Note that this formula is relative and should be applied after  an initial refractive index value is assigned to the starting phase. To this end, the 
average refractive index in the drop in the first image is assumed to be equal to the refractive index of the initial mixture. The full transformation of refractive index into concentration then follows from a calibration curve of the refractive index  of the dispersion $n$ as a function of the colloid concentration $\varphi$. Such a curve was obtained independently using an Abbe refractometer (Atago DR-A1, wavelength $589$~nm), and is displayed in Fig.~\ref{fig:FlowCurve}b. These data are superimposed with other measurements performed using other commercial grades: Ludox TM (wavelength $589$~nm) and Ludox TMA~\cite{Ning:08} (wavelength $632.8$~nm), as well as autocalibration curves from $10$ of our experiments  (wavelength $632.8$~nm). Indeed, from the image processing, one can extract the average refractive index $<n>$ as a function of the average colloid concentration $<\varphi>$ in the drop during the drying. All these data are nicely fitted by the linear law $n = 0.145\varphi + 1.3362$. 
 Note that silica and water both display a very low chromatic dispersion ($\delta n / n < 0.1\%$ between $589$ and $632.8$~nm) preventing  us to observe a difference between the two wavelengths we used. 
This calibration step finally leads to Fig~\ref{fig:Method}e. The colloid concentration field is also shown in polar coordinates below Fig~\ref{fig:Method}e to highlight the axisymmetric nature of the evaporation when the meniscus freely recedes. Hence, the available 2D information can be further compressed and averaged azimutally so as to obtain the concentration profiles shown in the results section.

Mach--Zehnder interferometry is in principle an optical technique of incredible precision. An accuracy of $2\pi/20$ on the phase $\phi$ is classically mentioned (or even sometimes better), which  corresponds in our case to an accuracy of $\simeq1.5\times10^{-3}$ on the colloid concentration $\varphi$. Even if the high accuracy of the technique results in extracting very smooth colloid concentration profiles allowing us to perform accurate spatial derivatives (see Section~\ref{sec:Results}), we do not expect to have achieved such an absolute accuracy because of several inevitable errors inherent in our set-up and our experiments (slight evaporation during the deposit of the drop, drifts, vibrations\dots). Hence, instead of the above-mentioned synthetic value of the accuracy, we statistically estimate it by repeating several times the same experiment (see the experiments presented in detail in Fig.~\ref{fig:Sequence_AS40normal4}--\ref{fig:ProfilFit}). The standard deviation provides us an absolute accuracy $< 0.5\%$ on the colloid concentration $\varphi$.

\section{Results \label{sec:Results}}

\subsection{Drying kinetics and observed phenomenology \label{sec:dryinkkinetics}}

\begin{figure}[htbp]
\begin{center}
\includegraphics{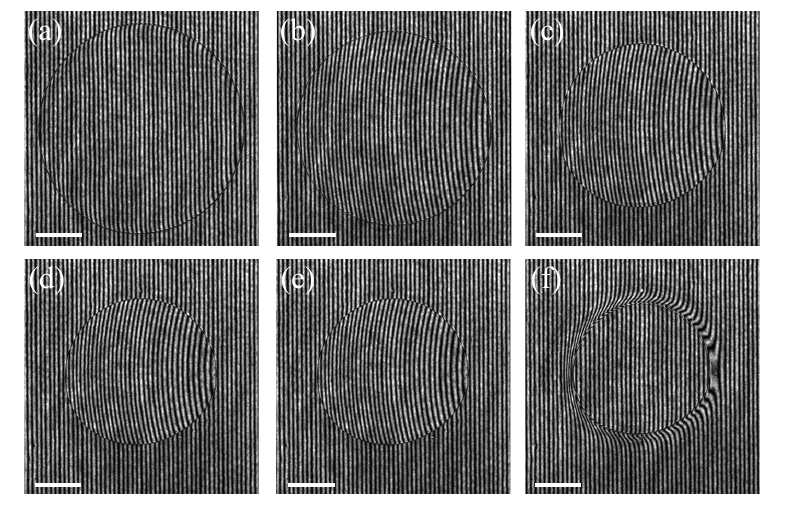}
\caption{(a)--(f) Typical interference field observation of the drying dynamics for $t=0$, $19.29$, $38.58$, $57.87$, $77.16$ and $96.45$~min. The scale bar is equal to $500~\mu$m and the drying conditions are $h=150~\mu$m, $\varphi_0 = 0.24$, $R_0 =1.04$~mm, $\tau_f =172$~min and $\text{Pe} = 5.3$ (see also Supplementary Movie 1).
\label{fig:Sequence_AS40normal4}}
\end{center}
\end{figure}

\begin{figure}[htbp]
\begin{center}
\includegraphics{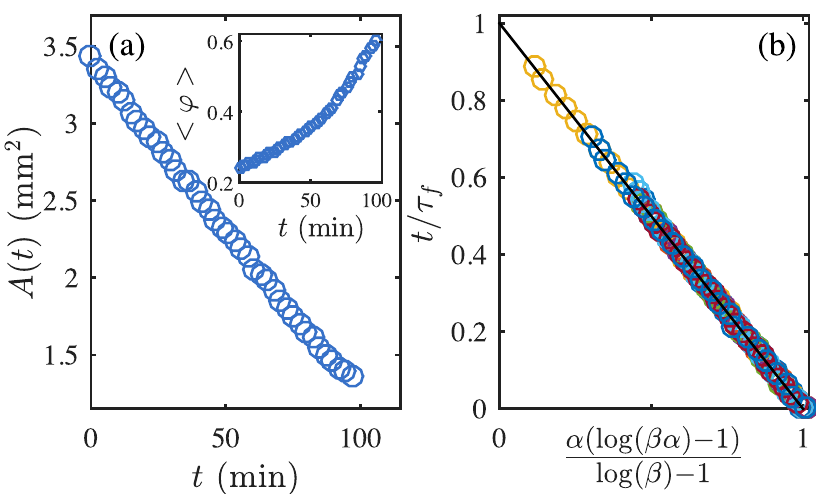}
\caption{(a) Typical evolution of the drop area $A$ vs.\ time $t$. Inset: corresponding average colloid concentration estimated from the colloid mass balance given in eqn~(\ref{eq:phimoy}). The drying conditions are those provided in Fig.~\ref{fig:Sequence_AS40normal4}. (b) Rescaled kinetics given by eqn~(\ref{eq:soldryindynamics}) for $10$ experiments performed in different conditions ($h=[150; 300]~\mu$m, $\varphi_0 = [0.05; 0.14; 0.24]$, $R_0 \simeq 1.0$--$1.7$~mm, $\tau_f \simeq 170$--$300$~min, and $\text{Pe} \simeq [5.3 ; 11.6]$).
\label{fig:Dynamics}}
\end{center}
\end{figure}

Figure~\ref{fig:Sequence_AS40normal4} shows a typical interferometric observation of the drying of the Ludox AS dispersion. Generally speaking, the drying drop remains cylindrical during most of the drying, except in the very late stage. In this experiment in which $h=150~\mu$m, $\varphi_0 = 0.24$, $R_0 =1.04$~mm, $T=21^{\circ}$C and $\text{R.H.} = 0.5$, the drop loses its axi-symmetricity  for $t\gtrsim75$~min. Actually, the meniscus first recedes freely due to solvent evaporation on the PDMS surface, but thin deposits are systematically observed on the wafers in the final stage, coupled with subsequent non axisymmetrical slight deformations of the drop. These deposits, clearly evidenced using bright field microscopy in ref.~\cite{Loussert:16,Bouchaudy:2019}, are certainly responsible for the deformations of the interference pattern observed outside the drop, at the vicinity of the meniscus, in the very late stage (see Fig.~\ref{fig:Sequence_AS40normal4}f).  
As explained in Section~\ref{sec:protocolinterf}, from a simple image analysis, one can extract from such pictures the corresponding temporal evolutions of the drop area $A(t)$ and of the average colloid concentration in the drop $<\varphi>$. These latter are displayed in Fig.~\ref{fig:Dynamics}a for our reference experiment. In all the experiments, $<\varphi>$ increases monotonically from $<\varphi> = \varphi_0$ up to $<\varphi> \simeq 0.6$ and the drop further delaminates from the wafers and cracks eventually form. Note that we are not interested in this work in such mechanical instabilities but we rather focus on the first stage of drying (corresponding to the concentration process up to the final consolidation), in particular when the drop remains perfectly cylindrical with no deposits.

The evolution of the drop area $A(t)$  as a function of time  appears nicely fitted by eqn~(\ref{eq:soldryindynamics}), leading to an experimental estimate of the characteristic drying time $\tau_f\simeq172~$min for this experiment. Figure~\ref{fig:Dynamics}b shows the dimensionless drying kinetics for $10$ experiments performed in different conditions, all collapsing on the theoretical prediction of eqn~(\ref{eq:soldryindynamics}). From all the fitted values of $\tau_f$ (knowing the respective relative humidity R.H. and temperature $T$), one can estimate the diffusivity of the water vapor in air using eqn~(\ref{eq:dryingtimes}). We find $D_s \simeq 2.6\pm 0.1 \times 10^{-5}$~m$^2$/s, in good agreement with the literature value (the uncertainty is the standard deviation calculated for the $10$ different experiments), validating 
the theoretical description of the drying kinetics presented in Section~\ref{sec:ConfinedDryingTheory}.

From the fitted values of $\tau_f$, one can also estimate the values of the P\'eclet number for our experiments using eqn~(\ref{eq:Pe}). As previously mentioned, $\text{Pe}\simeq 5.3-11.6$ which suggests the presence of gradients of colloid concentration during the drying. This is indeed what can be qualitatively observed from the typical interferograms presented in Fig.~\ref{fig:Sequence_AS40normal4}. Except at the initial and final stages of the concentration process for which the concentration field is quasi-homogeneous (see Fig.~\ref{fig:Sequence_AS40normal4}a with  $<\varphi> \simeq \varphi_0$ and Fig.~\ref{fig:Sequence_AS40normal4}f with $<\varphi> \simeq 0.6$), the interference patterns within the drop are clearly distorted indicating the presence of gradients of refractive index and therefore of concentration (see Fig.~\ref{fig:Sequence_AS40normal4}b--e). 

\subsection{Concentration fields in the drying drop}

From the temporal sequence of interferograms, one can extract the 2D colloid concentration fields $\varphi(r,\theta,t)$ following the procedure explained in Section~\ref{sec:protocolinterf}.
These 2D concentration fields are axisymmetrical, except in the late stage of drying when the concentration at the interface generally reaches $\varphi(R(t),t) \gtrsim 0.5$.
From these 2D axisymmetric maps, we then compute the angular-averaged concentration profiles $\varphi(r,t)$.

\subsubsection{Diffusive transport and collective diffusion coefficient}

\begin{figure}[htbp]
\begin{center}
\includegraphics{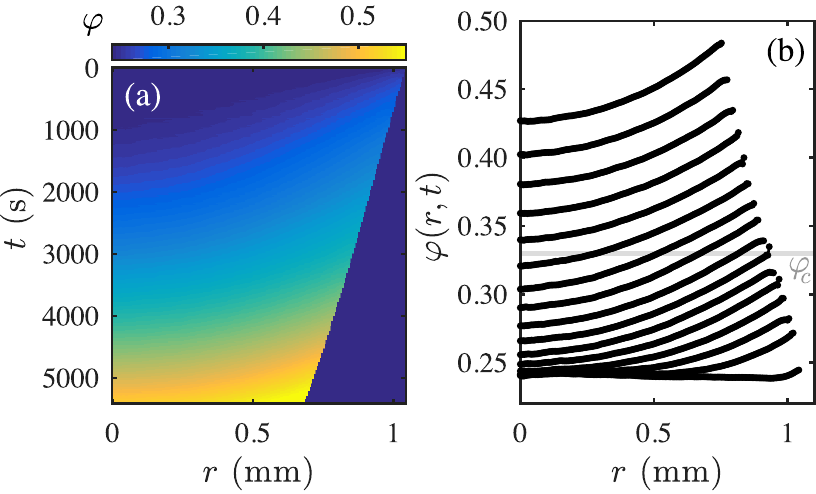}
\caption{(a) Space-time diagram of the angular-averaged concentration profiles $\varphi(r,t)$. (b) Same data at different times to highlight the build-up of concentration gradients. The time between two curves is $5~$min. The drying conditions are $h=150~\mu$m, $\varphi_0 = 0.24$, $R_0 =1.04$~mm, $\tau_f =172$~min and $\text{Pe} = 5.3$, identically to Fig.~\ref{fig:Sequence_AS40normal4} and~\ref{fig:Dynamics}a. See also the Supplementary Movies 2 and 3.
\label{fig:Profils}}
\end{center}
\end{figure}

Figure~\ref{fig:Profils} shows the temporal evolution of the colloid concentration profile within a drying drop of Ludox AS dispersion for the reference experiment involving $h=150~\mu$m, $\varphi_0\simeq0.24$, $R_0 =1.04$~mm, $\tau_f =172$~min and $\text{Pe} = 5.3$. In Fig.~\ref{fig:Profils}b, we plot only a few curves along the drying process for the sake of clarity. The time between two curves is $5$~min. These concentration profiles clearly reveal at short times the formation of a diffuse boundary layer at the receding meniscus, which gradually invades the entire drop radius. At longer time scales, the concentration gradients are rather weak along the drop, as previously reported by Loussert {\it et al.}\ on the same system~\cite{Loussert:16}.

\begin{figure}[htbp]
\begin{center}
\includegraphics{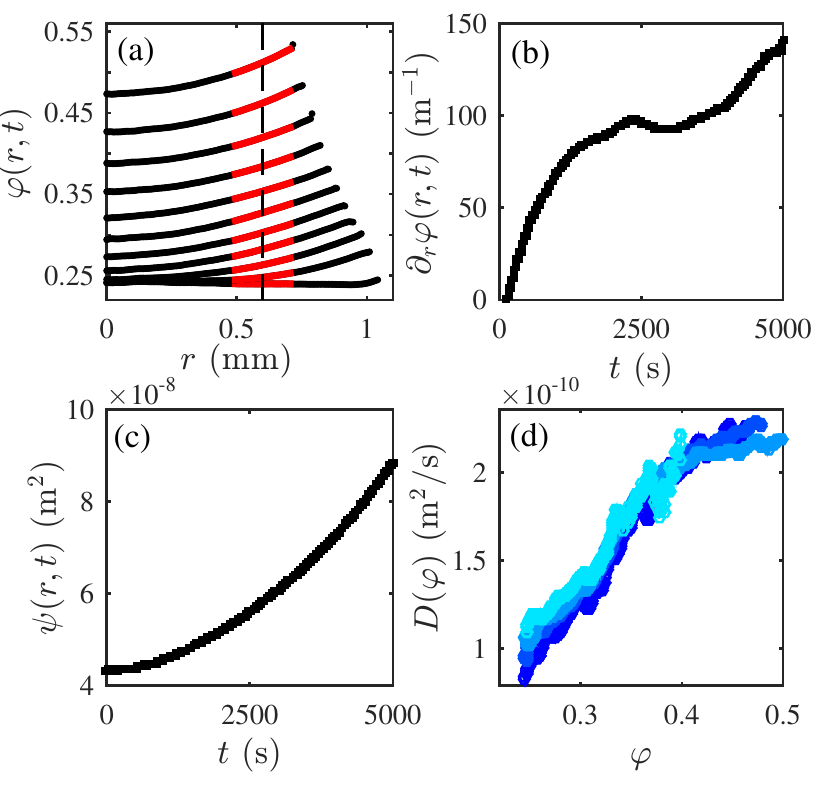}
\caption{(a) Concentration profiles at different times $t$, same data as in Fig.~\ref{fig:Profils}. The time between two curves is $500~$s. The red curves are local fits by a polynomial of degree 2 around $r = 0.6$~mm. 
(b) Temporal evolution of the gradient of the concentration profile at $r=0.6$~mm computed from the fits shown in (a). 
(c) $\psi(r,t)$ vs.\ $t$  for $r = 0.6$~mm, see eqn~(\ref{eq:defpsi}). (d) Estimates of $D(\varphi)$ using eqn~(\ref{eq:intemassbalance}) at different $r = [0.2; 0.4; 0.6; 0.8]$~mm (from dark to light blue).	
\label{fig:AnalyzProfils}}
\end{center}
\end{figure}

Assuming that the mass transport in the drying drop is well-described by diffusion only (see Section~\ref{sec:ConfinedDryingTheory} and in particular eqn~(\ref{eq:Transport}) and (\ref{BC1})), we can now extract the collective diffusion coefficient of the dispersion  from the measurements of $\varphi(r,t)$. To estimate  precise values of $D(\varphi)$, we proceed as follows, based on 
the global  solute conservation equation (obtained from the spatial integration of eqn~(\ref{eq:Transport})): 
\begin{eqnarray}
\frac{\partial \psi(r,t)}{\partial t } = r D(\varphi) \frac{\partial \varphi}{\partial r}\,, \label{eq:intemassbalance}
\end{eqnarray}
relating the diffusive flux at a given radius $r$ to the temporal variation of 
\begin{eqnarray}
\psi(r,t) = \int_0^r u \varphi(u,t) \text{d}u\,, \label{eq:defpsi}
\end{eqnarray}
for $r<R(t)$. 
All quantities in eqn~(\ref{eq:intemassbalance}) can be computed from the experimental measurements of $\varphi(r,t)$ at a given radial position $r$ in the drop, leading thus to values of $D$ for the concentration $\varphi$ estimated at $r$.

Figure~\ref{fig:AnalyzProfils} shows such a procedure for a given experiment. By considering the temporal evolution of $\varphi(r,t)$ at a fixed position $r$ in the drop (see Fig.~\ref{fig:AnalyzProfils}a), the concentration gradients at this position $r$ are estimated from local fits of the profiles by a polynomial of degree 2 (see Fig.~\ref{fig:AnalyzProfils}b), while $\psi(r,t)$ is computed using a numerical integration of the profiles according to eqn~(\ref{eq:defpsi}) (see Fig.~\ref{fig:AnalyzProfils}c). 
Then, at a time $t$ to which a specific concentration $\varphi(r,t)$ corresponds, $D$ can be deduced at a given $r$ from the measurements of $\partial_r \varphi$ and $\partial_t \psi$ using eqn~(\ref{eq:intemassbalance}), see Fig.~\ref{fig:AnalyzProfils}d. $\partial_t \psi$ is also obtained from local fits of $\psi(r,t)$ by a polynomial of degree 2.

Figure~\ref{fig:AnalyzProfils}d displays the mass transport coefficient $D(\varphi)$ measured at various radial positions $r$ in the drop for the experiment shown in Fig.~\ref{fig:Sequence_AS40normal4}, \ref{fig:Dynamics}a and \ref{fig:Profils}. Interestingly, all these measurements of $D(\varphi)$ obtained at various $r$ roughly collapse on a single curve suggesting that the mass transport in the drop is correctly described by eqn~(\ref{eq:Transport}), and thus governed by diffusion only. Therefore, the measured coefficient $D(\varphi)$  is effectively the collective diffusion coefficient of the dispersion.  From these curves, we observe that $D(\varphi)$ appears to be significantly larger that the Stokes-Einstein estimate $D_0\simeq2\times10^{-11}$~m$^2$/s, and increases monotonically from $D \simeq 1 \times 10^{-10}$~m$^2$/s at $\varphi \simeq 0.25$ up to $D \simeq 2.4 \times 10^{-10}$~m$^2$/s at $\varphi \simeq 0.5$.

\begin{figure}[htbp]
\begin{center}
\includegraphics{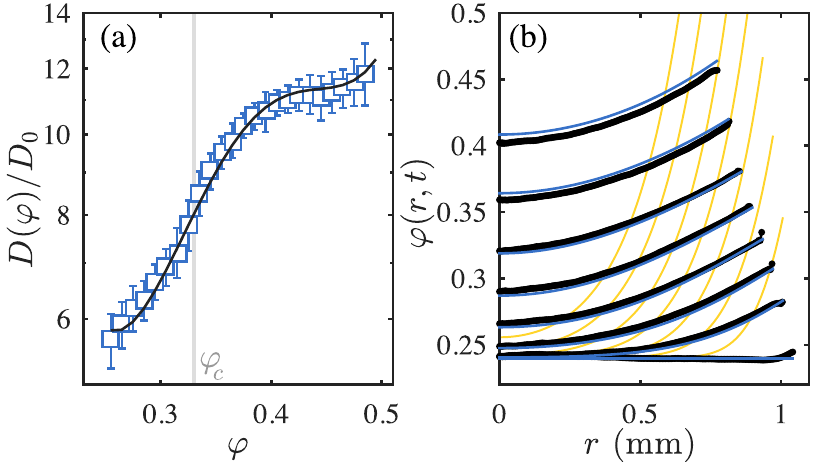}
\caption{(a) Normalized collective diffusion coefficient of the Ludox AS dispersion estimated from the measurements of $\varphi(r,t)$ using eqn~(\ref{eq:intemassbalance}), see text. The errorbars are calculated using the standard deviation over the equivalent of $15$ experimental curves of $D(\varphi)$ obtained with $h=150~\mu$m and $\varphi_0 = 0.24$. 
The black line is a fit of $D(\varphi)$.
(b) Concentration profiles already shown in Fig.~(\ref{fig:Profils}). The time between two curves is $10~$min here. The blue lines are the numerical solutions of eqn~(\ref{eq:Transport}) and (\ref{BC1}) using the fit for $D(\varphi)$ displayed in (a). The orange lines are the numerical solutions of the same equations for $D(\varphi) = D_0$.
 \label{fig:ProfilFit}}
\end{center}
\end{figure}

Figure~\ref{fig:ProfilFit}a shows the measured collective diffusion coefficient $D(\varphi)$ normalized by $D_0$. This curve was obtained from three experiments performed with $h=150~\mu$m and $\varphi_0 = 0.24$ (as well as $R_0=[1.04; 1.47; 1.71]$~mm, $\tau_f = [172; 300; 213]$~min, and $\text{Pe}=[5.3;6.1;11.6]$), for which $D(\varphi)$ was estimated at several positions $r$ in the drop.
Thus, $D(\varphi)$ and its errorbars plotted in Fig.~\ref{fig:ProfilFit}a are calculated from the mean value and the standard deviation over $15$ curves of $D(\varphi)$.
These results, and particularly the relatively small errorbars, clearly demonstrate that the colloid transport inside the drying drop is indeed well-described by the diffusion equation eqn~(\ref{eq:Transport}), with a collective diffusion coefficient $D(\varphi)$ given by the data displayed in Fig.~\ref{fig:ProfilFit}a. The measured $D(\varphi)$ for the investigated charged colloidal dispersion is indeed significantly larger than the Stokes-Einstein prediction $D_0$ since $D(\varphi)/D_0$ increases from about $5$ to $12$ in the range of $\varphi=0.24$--$0.5$. Such a result obviously reveals the importance of the colloidal 
interactions (missing in $D_0$) to describe the relaxation of concentration gradients.

Figure~\ref{fig:ProfilFit}b presents the temporal evolution of the concentration profiles of our reference experiment 
and the numerical solutions of  eqn~(\ref{eq:Transport}) and (\ref{BC1}) computed for these experimental  conditions using  either the fit of $D(\varphi)$ shown in Fig.~\ref{fig:ProfilFit}a or $D(\varphi)=D_0$ for the collective diffusion coefficient. Details about the numerical resolution of the theoretical model presented in Section~\ref{sec:ConfinedDryingTheory} are provided in Annex~A.
As expected, the solutions of the theoretical model using the measured  $D(\varphi)$ nicely predict the overall drying dynamics, while considering a constant coefficient equal to $D_0$ strongly overestimates the concentration gradients, and poorly predicts the concentration process. 
This comparison of the experimental data with the theoretical predictions of the model presented in Section~\ref{sec:ConfinedDryingTheory}
again confirms that the mass transport in the drop is well-described by a diffusion equation, but  with a collective diffusion coefficient $D(\varphi)$  significantly larger than $D_0$ for our charged colloidal dispersion.

\subsubsection{Evidence of the role of buoyancy-driven free convection \label{sec:roleofbuoyancy}}

Actually, free convection has been clearly evidenced in such confined drying experiments, and for many different binary mixtures~\cite{Selva:12,Lee:14,Pradhan2018,Daubersies:12}, including also the charge-stabilized dispersion studied here~\cite{Loussert:16,Bouchaudy:2019}. Indeed, as mentioned in Section~\ref{sec:ConfinedDryingTheory}, solute concentration gradients unavoidably generate buoyancy-driven free convection in the drop, as these gradients are orthogonal to the gravity. Considering a constant viscosity $\eta$ over the drop and the lubrication approximation, i.e.~concentration gradients  on a length scale much larger than the cell height $h$, one can easily derive that the density gradient naturally drives free convection whose maximal velocity scales as~\cite{Selva:12}:
\begin{eqnarray}
v_m   \sim \frac{g (\rho_{c}-\rho_{s}) h^3}{\eta} \frac{\delta \varphi}{R}\,, \label{eq:vmscaling}
\end{eqnarray}
where $\delta \varphi$ is the typical variation in concentration along the drop radius $R$, $\rho_{c}$ the density of the colloids and $\rho_{s}$ that of the solvent. In the case of the Ludox AS dispersion studied here, Loussert {\it et al.}\ indeed measured such radial recirculating flows using fluorescent tracers added in the dispersion. They observed that these flows vanish when the local concentration reaches $\varphi \simeq 0.32$, in agreement with the divergence of the viscosity $\eta$ revealed in Fig.~\ref{fig:FlowCurve}a. They also measured using particle tracking velocimetry, the strong dependence of $v_m$ with the cell thickness $h$, in accordance with the scaling of eqn~(\ref{eq:vmscaling})~\cite{Loussert:16}.  Loussert {\it et al.}\ nevertheless claimed that free convection had a negligible impact on their measurements~\cite{Selva:12}. However, a crude estimate of the impact of free convection on the diffusive transport can be made with the following dimensionless number: 
\begin{eqnarray}
\frac{v_m  h}{D} =  \frac{h}{R} \frac{g (\rho_s-\rho_w) h^3 \delta \varphi}{\eta D} = \frac{h}{R}\text{Ra}\,,\label{eq:RaScaling}
\end{eqnarray} 
where $\text{Ra}$ is the Rayleigh number. In the dilute regime, in which $\eta \simeq 15$~mPa.s and $D \simeq 10^{-10}$~m$^2$/s, one finds $(h/R)\text{Ra} \simeq 100 \gg 1$ (even for $h=150~\mu$m, $\delta \varphi \simeq 0.025$, and  $R=1$~mm) suggesting that natural convection should always dominates mass transport in our experiments.  Going beyond this simple scaling analysis and the conclusion drawn by Loussert {\it et al.}~\cite{Loussert:16}, we demonstrate herebelow that free convection actually influences	the mass transport within the drop, but only for large cell height, typically $h \geq 200~\mu$m for the Ludox AS dispersion at $\varphi_0 = 0.24$.

\begin{figure}[htbp]
\begin{center}
\includegraphics{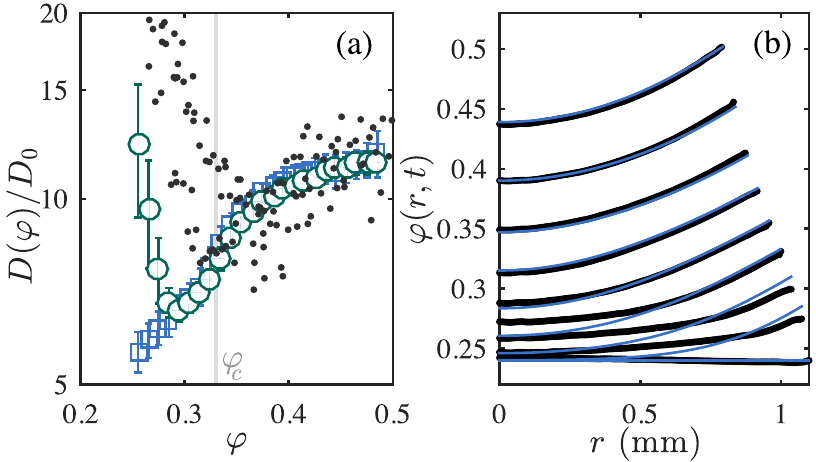}
\caption{(a) Measurements of the normalized collective diffusion coefficient for $h=300~\mu$m (\textcolor{green2}{$\circ$}) and $h=150~\mu$m (\textcolor{blue}{$\square$}, same data shown in Fig.~\ref{fig:ProfilFit}). The errorbars of $D(\varphi)$ for $h=300~\mu$m are calculated using the standard deviation over the equivalent of $15$ experimental curves of $D(\varphi)$ obtained with $\varphi_0 = 0.24$, see text. Data from~\citet{Loussert:16} for the same system are also plotted ($\bullet$). 
(b) Temporal evolution of the concentration profiles for $h=300~\mu$m. The other drying conditions are $\varphi_0 = 0.24$, $R_0 = 1.11$~mm, $\tau_f = 178$~min, and $\text{Pe}=5.8$. The time between two curves is $10$~min. The blue lines are the theoretical predictions using eqn~(\ref{eq:Transport}) and (\ref{BC1}) and $D(\varphi)$ given by the fit shown in Fig.~\ref{fig:ProfilFit}a, i.e.\ considering a purely diffusive mass transport within the drying drop.
 \label{fig:DiffHFit}}
\end{center}
\end{figure}

To reveal the role of free convection on the colloid concentration dynamics, we performed similar experiments and analysis as above, but for a larger cell height, namely $h=300~\mu$m. While the phenomenology observed is similar to the one reported in Section~\ref{sec:dryinkkinetics} and the drying kinetics still follows eqn~(\ref{eq:soldryindynamics}) (see Fig.~\ref{fig:Dynamics}b), the measurements of $D(\varphi)$ from the concentration profiles using eqn~(\ref{eq:intemassbalance}) now differ significantly from the case $h=150~\mu$m, as shown in Fig.~\ref{fig:DiffHFit}a, at least in the liquid regime. 

More precisely, the measured values of $D(\varphi)$ with $h=300~\mu$m are now significantly higher than those measured with $h=150~\mu$m in the range of $\varphi = 0.24$--$0.3$. Moreover, the errorbars 
are also very large in this concentration range. Strikingly, the values of $D(\varphi)$ measured for this set of experiments collapse at higher concentrations, when $\varphi > 0.3$, on the data obtained with a cell height $h=150~\mu$m. 
Note that this new curve was obtained from three experiments performed with $h=300~\mu$m and $\varphi_0 = 0.24$ (as well as $R_0=[1.11; 1.15; 1.24]$~mm, $\tau_f = [178; 207; 208]$~min, and $\text{Pe}=[5.87; 5.38; 6.24]$), for which $D(\varphi)$ was again estimated at several positions $r$ in the drop.  
Thus, $D(\varphi)$ and its errorbars plotted in Fig.~\ref{fig:DiffHFit}a are calculated from the mean value and the standard deviation over $15$ curves of $D(\varphi)$. It is worth mentioning that such large errorbars are not only the results of a dispersion of data between the experiments, but are also obtained between the different positions $r$ probed within the drop at each experiment.

These results, and in particular the large errorbars, suggest that the dynamics of concentration for a height $h=300~\mu$m is not described in the liquid regime (when $\varphi< 0.3$) by eqn~(\ref{eq:Transport}), i.e.\ by diffusion only.
Figure~\ref{fig:DiffHFit}b, which displays the temporal evolution of the concentration profiles for an experiment with $h=300~\mu$m, along with the numerical solution of the purely diffusive model for the colloid transport (i.e.\ eqn~(\ref{eq:Transport}) and (\ref{BC1}) with $D(\varphi)$ given by the data shown in Fig.~\ref{fig:ProfilFit}a) for these experimental conditions, confirms this result. The experimental concentration profiles are not correctly fitted by the numerical solution at early times scales (i.e.\ when $\varphi(r,t) < 0.3$), whereas the latter well-describes the whole profiles at longer time scales. All this suggests that buoyancy-driven free convection in the drop affects the transport process, obviously as long as the dispersion is in a liquid state.

Evaluating the impact of free convection on the solute mass transport in our configuration would require a complete theoretical description of the drying process, coupling the Stokes equation (including buoyancy)  with the transport equation eqn~(\ref{eq:TransportConv}).
This task is clearly beyond the scope of the present work and would demand careful numerical resolution, in particular due to the 
strong variation of the viscosity with $\varphi$ (see Fig.~\ref{fig:FlowCurve}a). Salmon and Doumenc recently investigated theoretically the free convection in a drying binary mixture (solvent $+$ non-volatile solute), assuming constant viscosity $\eta$ and diffusion coefficient $D$~\cite{Salmon2019}. While the authors focused on a uni-directional configuration, they mentioned that their 
theoretical derivation also applies to our specific 2D geometry. 
Authors of ref.~\cite{Salmon2019} demonstrated using a standard Taylor-like approach~\cite{Young1991}  that the solute concentration profile  remains described by a diffusion equation: 
	\begin{eqnarray}
	&&\frac{\partial \varphi}{\partial t} = \frac{1}{r}\frac{\partial }{\partial r} \left(r  D_{\text{eff}} \frac{\partial \varphi}{\partial r}  \right)\,,
	\end{eqnarray}
but with an effective dispersion coefficient, taking into account both diffusion and free convection on the solute mass transport, given by
\begin{eqnarray}
 D_{\text{eff}} = D\left[1 + \frac{1}{362880}\left(\frac{g (\rho_s-\rho_w) h^4}{\eta D}\frac{\partial \varphi}{\partial r}\right)^2 \right] \,. \label{eq:Deffdim}
\end{eqnarray} 
According to eqn~(\ref{eq:Deffdim}), 
free convection  does not impact diffusion when $D_{\text{eff}} \simeq D$, and thus for:
\begin{eqnarray}
\begin{split}
\frac{ D_{\text{eff}}}{D}-1 &= \frac{1}{362880}\left(\frac{g (\rho_s-\rho_w) h^4}{\eta D}\frac{\partial \varphi}{\partial r}\right)^2 \\
&\simeq  \frac{1}{362880}\left(\frac{h}{R}\text{Ra}\right)^2 \ll 1\,.
\end{split} 
\label{eq:Deffdim2} 
\end{eqnarray} 
A similar inequality can be also derived using scaling arguments~\cite{Gu2018},  see  eqn~(\ref{eq:RaScaling}),   but without the  non intuitive  numerical prefactor $362880$. 

\begin{figure}[htbp]
\begin{center}
\includegraphics{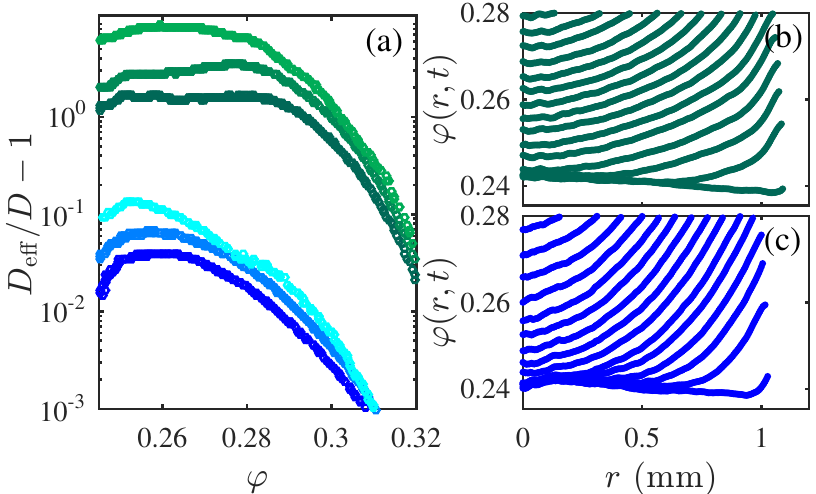}
\caption{(a) Reduced effective diffusivity estimated using eqn~(\ref{eq:Deffdim}) for two similar experiments ($\varphi_0 = 0.24$), but performed for two cell heights: $h=150~\mu$m (blue, $R_0=1.04$~mm, $\tau_f=172~$min, and $\text{Pe}=5.3$) and $h=300~\mu$m (green, $R_0=1.11$~mm, $\tau_f=178~$min, and $\text{Pe}=5.8$). $D_{\text{eff}}$ is computed from the gradient of the concentration profiles $\varphi(r,t)$ at $r=500$, $700$, and $900~\mu$m in the drops (from dark to light colors).  
The viscosity and the collective diffusion coefficient of the dispersion in this concentration range were estimated using the fits shown in Fig.~\ref{fig:FlowCurve}a and Fig.~\ref{fig:ProfilFit}a, respectively. (b)-(c) Corresponding concentration profiles $\varphi(r,t)$ at small time scales, for $h=300~\mu$m (b) and $h=150~\mu$m (c). The time between two curves is $150~$s. \label{fig:Buoyancy}}
\end{center}
\end{figure}

Even if our experimental configuration is more complex than the model case treated in ref.~\cite{Salmon2019}, in particular due to the variations of $\eta$ and $D$ with the colloid concentration $\varphi$, we will assume that eqn~(\ref{eq:Deffdim}) enables, as a first step, to evaluate the impact of buoyancy on mass transport in our case. 
Thus, Fig.~\ref{fig:Buoyancy}a shows the reduced coefficient $ D_{\text{eff}}/D-1$ at the early stages of the drying  
(i.e.\ when $\varphi < 0.3$) for two experiments performed in similar conditions, but for two different cell heights $h$ : $150~\mu$m and $300~\mu$m. To plot this, the viscosity $\eta$ and collective diffusion coefficient $D$ in eqn~(\ref{eq:Deffdim}) were estimated using the fits shown in Fig.~\ref{fig:FlowCurve}a and Fig.~\ref{fig:ProfilFit}a, while the concentration gradients were calculated using local fits of the concentration profiles for various positions $r$ in the drop, as shown in Fig.~\ref{fig:AnalyzProfils}a.
In the case $h=150~\mu$m, the values of the reduced coefficient range initially from $10^{-2}$ to $10^{-1}$, and rapidly drops below $10^{-2}$ for $\varphi>0.29$, mainly due to the increase of the viscosity with $\varphi$ ($D$ remains almost constant in this concentration range). In the case $h=300~\mu$m, $ D_{\text{eff}}/D-1 > 1$  
when $\varphi< 0.29$, before decreasing below $10^{-1}$  for $\varphi \geq 0.31$, again due to the strong increase of $\eta$ with $\varphi$.  
Hence, without intending to claim that the theoretical description given in ref.~\cite{Salmon2019} strictly applies to our experimental case because of the variations of  $D$ and $\eta$ with $\varphi$, the inequality given in eqn~(\ref{eq:Deffdim2}) combined with our measurements clearly suggest that buoyancy-driven free convection does not dominate colloid mass transport for $h=150~\mu$m, whereas it impacts mass transport when $h=300~\mu$m for moderate viscosity ($\varphi< 0.29$). 
This rationalizes our measurements of $D(\varphi)$ reported in Fig.~\ref{fig:DiffHFit}a, about both the collapse of the data measured with $h=300~\mu$m with those measured at $h=150~\mu$m for $\varphi>0.3$, and the measurement of a higher effective dispersion coefficient with $h=300~\mu$m than the collective diffusion coefficient for $\varphi<0.3$, convection increasing mass transfer. The latter is also noticeable when looking at the concentration profiles at early time scales, Fig.~\ref{fig:Buoyancy}b obtained for $h=300~\mu$m evidencing a faster growth of the diffuse layer in the drop as compared to Fig.~\ref{fig:Buoyancy}c obtained for $h=150~\mu$m (i.e.\ with purely diffusive transport). Finally, note that while free convection appears to  impact significantly the transport of colloids for $\varphi < 0.3$ in the case $h=300~\mu$m, the vertical concentration gradients expected to develop along the thickness $h$ of the drop~\cite{Salmon2019} cannot be quantified by our interferometric technique which only measures an average concentration over $h$.

\subsection{Comparison with the data of Loussert {\it et al.} \label{sec:Loussert}}
Loussert {\it et al.}\ also measured $D(\varphi)$ over the same concentration range and for the same system, but using Raman confocal spectroscopy scans along the drop diameter~\cite{Loussert:16}. Their data are plotted in Fig.~\ref{fig:ProfilFit}a, along with our measurements at $h=150$ and $300~\mu$m. Despite their wide dispersion, the data of Loussert {\it et al.}\ converge towards our measurements at high concentration, but a significant
increase is systematically reported at low volume fractions, when $\varphi<0.35$.
It is worth noting that some of these data were obtained with a cell height $h=250~\mu$m, for which free convection strongly influences colloid mass transport, thus leading to larger values of the diffusivity. Nevertheless,
some of these data were also obtained  with a cell height $h=170~\mu$m, for which free convection is expected to have a negligible influence. Actually, Loussert {\it et al.}\ estimated $D(\varphi)$ using the boundary condition eqn~(\ref{BC1}) only,  by computing the concentration gradient at the receding meniscus while measuring simultaneously the drying rate. Their setup was in particular not resolved enough (spatially $\simeq 30~\mu$m, and temporally $\simeq 0.5$~profile/min) and not accurate enough ($\pm 2\%$) to capture precisely the growth of the diffuse boundary layer that invades the drop at small time scales, i.e.\ in the concentration range $\varphi \simeq 0.24$--$0.3$ (see Fig.~\ref{fig:Buoyancy}c). In addition, to avoid numerical differentiation of $\varphi(r,t)$ at the receding meniscus due to the high uncertainties of their measurements, they actually fitted their  concentration profiles using parabolas over the whole diameter, therefore assuming weak gradients and thus implicitly large values of $D$~\cite{Loussert:16}.

\section{Discussions and conclusions \label{sec:discconc}}

In the present work, we used Mach-Zehnder interferometry to probe mass transport in a drying drop of a charged colloidal dispersion. 
By coupling this optical technique with a 2D confined geometry, our setup leads to accurate measurements of the concentration profiles, from which we extracted precise values of the collective diffusion coefficient of the dispersion $D$ over a wide concentration range $\varphi = 0.24$--$0.5$, i.e.\ from the liquid dispersed state to the solid glass regime. Importantly, our experiments clearly evidenced that, when performed in a thin cell ($h=150~\mu$m), the mass transport in the drop is governed by diffusion, hence fully validating for the first time the theoretical description presented in Section~\ref{sec:ConfinedDryingTheory}, whereas when performed in a thicker cell ($h=300~\mu$m), it is impacted by the buoyancy free-convection, at least in the liquid dispersed regime (i.e.\ when $\varphi<0.3$).

The main differences between the Mach-Zehnder interferometry used in the present work and the techniques previously used to measure colloid concentration fields in drying dispersions~\cite{Loussert:16,Goehring2017} are its high spatial and temporal resolutions, $\simeq 5~\mu$m and $1$~profile/s, as well as its high concentration accuracy $\simeq 0.5\%$. Moreover, our techniques make it possible to obtain 2D concentration fields $\varphi(r,\theta)$ from which we can state that these are axisymmetrical, unlike the measurements made by Loussert {\it et al.}\ for instance, only obtained from scans along the drop diameter and assuming axisymmetry. Note that any dust on the PDMS coatings may pin, even temporarily, the receding meniscus at some locations, thus resulting in a drop asymmetry. All these advantages make interferometry a powerful tool to probe mass transport and allow us, among others, to reveal the growth of a diffusive layer at short times that progressively invades  the drop, 
and above all to measure the collective diffusion coefficient of the dispersion $D(\varphi)$ with a high  accuracy.

Our precise measurements of $D(\varphi)$ (see Fig.~\ref{fig:ProfilFit}a) clearly reveal the crucial role played by the colloidal interactions on collective diffusion, as $D(\varphi) \simeq 5$--$12 D_0$ over the concentration range $0.24$--$0.5$. The collective diffusion coefficient appears to monotonically increase with the concentration, and do not display any specific signature around the glass transition $\varphi \simeq 0.33$, despite the huge increase of the viscosity (see Fig.~\ref{fig:FlowCurve}a), and the appearance of a yield stress and a finite shear modulus at larger concentrations.

The values of $D(\varphi)$ being significantly larger than $D_0$, our measurements evidence that mass diffusion in such dispersions can not be modeled 
on the basis of isolated-particle theories. 
When considering inter-particle interactions, the collective diffusion coefficient $D(\varphi)$ of a  charge-stabilized dispersion   is generally expressed as~\cite{ROA2016} 
\begin{eqnarray}
D(\varphi) = D_0 K(\varphi)\left(\frac{\partial (\varphi z(\varphi))}{\partial \varphi}\right)_{T,\text{res}}\,, \label{eq:DDonnan}
\end{eqnarray}
where $K(\varphi)$ is the long-time sedimentation coefficient which accounts for the hydrodynamic friction of the solvent flow relatively to the colloids, 
and $z(\varphi)$ the compressibility factor which accounts for the colloidal interactions. $z(\varphi)$ is
related to the osmotic pressure $\Pi(\varphi)$ of the dispersion through
$\Pi(\varphi) V_c = k_B T\varphi z(\varphi)$, with $V_c$ the volume of a colloid~\cite{Nagele,Russel}. 
In eqn~(\ref{eq:DDonnan}), the dispersion is considered in equilibrium with a reservoir of known salinity (the so-called Donnan equilibrium), and 
the derivative is taken for fixed temperature $T$ and reservoir conditions.
Considering particles as neutral hard-spheres HS, i.e.\ including  $z^{\rm{HS}}(\varphi)$  and  $K^{\rm{HS}}(\varphi)$ in eqn~(\ref{eq:DDonnan}) (see e.g.~ref.~\cite{Peppin:06}) 
does not explain our data, as such a theoretical prediction is relatively close to $D_0$ in our concentration range  ($D^{\rm{HS}}(\varphi)\simeq 0.7$--$1.1 D_0$ for $\varphi<0.5$). Goehring {\it et al.}, who also reported measurements of $D$ for similar charge-stabilized dispersions (Ludox SM30, HS40, and TM50) but using a vertical 1D drying cell~\cite{Goehring2017},
also attempted to model their measurements  using eqn~(\ref{eq:DDonnan}). More precisely, they  additionnally included into $z(\varphi)$  the long-range electrostatic repulsions modeled using the Poisson-Boltzmann cell model fitted on  equations of states $\Pi$ vs.\ $\varphi$ measured in Donnan equilibrium for several reservoir salinities, while
still retaining the hard-sphere sedimentation coefficient $K^{\rm{HS}}(\varphi)$. 
Such a model overestimates greatly both their measurements and our data, particularly in the dilute regime ($\varphi \leq 0.3$) for which 
values as high as $D(\varphi) \simeq 40$--$100 D_0$ are predicted depending on the ionic strength of the dispersion. 
Because the measured equations of states  $\Pi$ vs.\ $\varphi$ are in excellent agreement with the cell model, particularly for low salinities (see ref.~\cite{Hallez2017}), the discrepancy between the 
measured $D(\varphi)$ and the predictions given by eqn~(\ref{eq:DDonnan}) thus suggests that the long-range electrostatic repulsion should also play a role on the 
sedimentation coefficient. 
Roa~{\it et al.}\ derived  the same  theoretical model in the context of the ultrafiltration of charge-stabilized dispersions at low salinity~\cite{ROA2016}, 
including however the  structure of a charged dispersion with long-range electrostatic repulsions into the  sedimentation coefficient~\cite{Banchio2008}.
 It should be noted that we do not know whether the model used in~ref.~\cite{ROA2016} to estimate $K(\varphi)$ applies to  the range of concentrations we have explored, $\varphi = 0.24$--$0.5$.  Note also that the above models  based on eqn~(\ref{eq:DDonnan})   are 
expected to describe correctly the dynamics of a liquid dispersion in Donnan equilibrium with a reservoir of known salinity, but may not strictly apply to a drying dispersion as the dynamics of the ionic species may also play a role. Moreover, 
the liquid to solid transition could  also introduce some additional complexity, as proposed for instance in ref.~\cite{Style:11} using linear poroelasticity (see also ref.~\cite{Bouchaudy:2019} for discussions, and ref.~\cite{MacMinn2016} for the case of non-linear poroelasticity), 
as well as the polydispersity of our dispersion of small colloids. 
 We therefore hope that the present study will further motivate the development of models that take into account the complexity of the physical interactions that occur in these systems, particularly at high concentration.

To further test such models, it would be useful to measure $D(\varphi)$ over the full concentration range $\varphi \simeq 0$--$0.6$. 
Actually, we performed similar experiments as explained above, but with dilute dispersions, namely $\varphi_0 \simeq 0.04$ and $\varphi_0 \simeq 0.14$ (with $h=150~\mu$m, data not shown). As for the case $h=300~\mu$m and $\varphi_0 \simeq 0.24$, the estimated values of $D$ at small concentrations are associated to large errorbars suggesting that mass transport is also impacted by free convection. This can be again rationalized using the reduced effective dispersion coefficient given by eqn~(\ref{eq:Deffdim2}). To go further, eqn~(\ref{eq:Deffdim2}) combined  with the viscosity values shown in Fig.~\ref{fig:FlowCurve}a can be used to estimate the minimal height $h$ necessary to 
avoid free convection in such measurements. The numerical resolution of eqn~(\ref{eq:Transport}) and (\ref{BC1}), assuming a diffusive transport with $D \simeq 1$--$10 D_0$ in the concentration range $\varphi = 0.04$--$0.24$, shows that measurements of $D(\varphi)$ would be possible (i.e.\ negligible impact of free convection) only for $h \leq 30$--$40~\mu$m. We performed several attempts of such confinements, but unfortunately, significant non axisymmetrical deformations of the drop always occurred, even in the liquid regime, preventing an accurate measure of $D$. Combining Mach-Zehnder interferometry with a vertical 1D directional drying cell as in~ref.~\cite{Goehring2017}, would probably be the best way to probe collective diffusion with a high accuracy and over the full concentration range, while avoiding the complexity due to free convection.
 In addition, as the colloidal interactions in  dispersions of silica nanoparticles can be tuned by playing on the pH or the ionic strength, it would also be very relevant to study their role on collective diffusion. Such perspectives would continue to improve the understanding of mass transport in colloidal dispersions, so we hope that our work will also motivate further experimental studies on this topic.

\section*{Conflicts of interest}
There are no conflicts to declare.

\section*{Acknowledgements}
The authors warmly thank G. Ovarlez for his help on the rheological measurements, and Y. Hallez  for the many valuable discussions. This work was financially supported by the Fonds de la Recherche Scientifique -- F.N.R.S. (Postdoctoral Researcher Position of BS).
SD is grateful for the financial support from the European Space Agency (ESA) and the Belgian Federal Science Policy Office (BELSPO) through PRODEX Evaporation.

\section*{Appendix A. Dimensionless model and numerical resolution}

The theoretical model of the confined drying of a drop of colloidal dispersion presented in Section~\ref{sec:ConfinedDryingTheory} is here presented in dimensionless form. Using the dimensionless variables $\tilde{t} = t/\tau_f$, $\tilde{r} =r/R_0$, $\tilde{R} =R/R_0$, and defining $\hat{D}(\varphi)=D(\varphi)/D_0$,  the equations related to the colloidal distribution inside the drop (eqn~(\ref{eq:Transport}) and (\ref{BC1})) write: 
\begin{eqnarray}
&&\frac{\partial \varphi}{\partial \tilde{t} }= \frac{1}{r}\frac{\partial}{\partial \tilde{r}} \left( \tilde{r} \frac{\hat{D}(\varphi)}{\text{Pe}} \frac{\partial \varphi}{\partial \tilde{r}}\right)\,,\label{eq:Transport_adi}\\ 
&& \left.\frac{\hat{D}(\varphi)}{\text{Pe}} \frac{\partial \varphi}{\partial \tilde{r}} \right|_{\tilde{r}=\tilde{R},\tilde{t}} = -\varphi(\tilde{r}=\tilde{R},\tilde{t})\frac{\text{d}\tilde{R} }{\text{d}\tilde{t}}\,, \label{BC1_adi} 
\end{eqnarray}
while the one concerning the drying kinetics, eqn~(\ref{eq:dryindynamics}), expresses:
\begin{eqnarray}
\frac{\text{d}\alpha}{\text{d}\tilde{t}} =\frac{1-\ln(\beta)}{\ln(\beta\alpha)}\,. \label{eq:soldryindynamics_adi}
\end{eqnarray}

To numerically solve eqn~(\ref{eq:Transport_adi}) and (\ref{BC1_adi}), we performed the following change of variable $\varphi(\tilde{r},\tilde{t}) \to \varphi(u,\tilde{t})$
with $u = \tilde{r} / \sqrt{\alpha}$ ranging on fixed coordinates $[0-1]$.
Using these variables, eqn~(\ref{eq:Transport_adi}) and (\ref{BC1_adi}) read:
\begin{eqnarray}
&&\frac{\partial (\alpha \varphi)}{\partial \tilde{t} }= \frac{1}{u}\frac{\partial}{\partial u} \left( u \left[\frac{\hat{D}(\varphi)}{\text{Pe}} \frac{\partial \varphi}{\partial u} + \frac{u}{2} \varphi\frac{\text{d}\alpha}{\text{d}\tilde{t}}\right]\right)\,,\label{eq:Transport_adi2}\\ 
&& \left[\frac{\hat{D}(\varphi)}{\text{Pe}} \frac{\partial \varphi}{\partial u} + \frac{u}{2} \varphi\frac{\text{d}\alpha}{\text{d}\tilde{t}}\right]_{u=1,\tilde{t}} = 0\,. \label{BC1_adi2} 
\end{eqnarray}
The solution of these two equations are computed with the solver of partial differential equations of Matlab ({\it pdepe}), along with the analytical solution of eqn~(\ref{eq:soldryindynamics_adi}), and the initial condition:
\begin{eqnarray}
\varphi(u,\tilde{t}=0) = \varphi_0\,.
\end{eqnarray}

\section*{Appendix B. Rheological measurements}

A rheological characterization of the colloidal dispersion Ludox AS was performed using a stress-controlled rheometer (Kinexus ultra$+$, Malvern) at $T = 22^\circ$C, and using various shear geometries: a double Couette cell for the samples of low viscosity and a striated cone--plate geometry for the sample of high viscosity. While samples with a volume fraction $\varphi<0.24$ were obtained by mass dilution of the stock dispersion with deionised water, those of higher volume fractions were obtained by slow evaporation of macroscopic volumes of the stock dispersion ($\simeq 20~$mL) under gentle stirring at $T \simeq 30^\circ$C and for several days. 
Using this latter method, we obtained two 
samples with $\varphi>0.24$, namely $\varphi \simeq 0.29$ and $0.32$ (from dry extract measurements), that look macroscopically homogeneous. The most concentrated sample, although extremely viscous, does not display any significant yield stress since the air--dispersion interface in the vial always relaxes towards a flat surface over long periods of time (the hydrostatic pressure associated with a water height of $1$~mm is $10$~Pa). 

\begin{figure}[htbp]
\begin{center}
\includegraphics{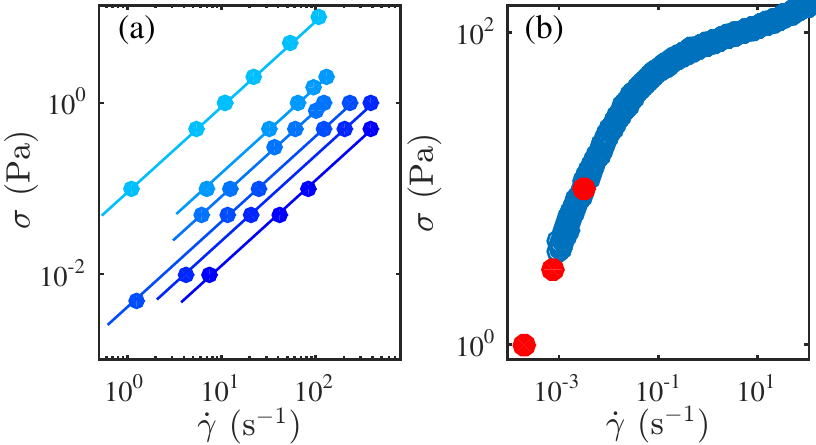}
\caption{(a) Steady flow curves of the shear stress $\sigma$ as a function of the shear rate $\dot{\gamma}$ for samples of dispersion of different volume fraction with $\varphi \leq 0.3$. The lighter the blue, the more concentrated the dispersion. While dots are experimental points obtained at imposed shear stresses, the continuous lines are fits of these data by $\sigma= \eta(\varphi) \dot{\gamma}$.
(b) Flow curve of $\sigma$ vs.\ $\dot{\gamma}$ for the sample of dispersion with $\varphi \simeq 0.32$. While the blue dots are obtained using a ramp of imposed shear rates, the red dots are steady measurements at imposed shear rates.
The low shear viscosity tends to $\eta \simeq 6 \pm 2$~Pa.s.
 \label{fig:RheologySI.pdf}}
\end{center}
\end{figure}

Figure~\ref{fig:RheologySI.pdf}a displays the shear stress $\sigma$ as a function of the shear rate $\dot{\gamma}$, obtained at imposed shear stress, for the samples of dispersion with $\varphi \leq 0.3$.
At these volume fractions, the dispersion clearly exhibits a Newtonian behavior leading to precise measurements of the viscosity $\eta$ as a function of the volume fraction $\varphi$. 
However, the sample at $\varphi \simeq 0.32$ displays a non-Newtonian behavior, as revealed by the ramp of imposed shear rates shown in Fig.~\ref{fig:RheologySI.pdf}b (ramp duration: $2$~min, pre-shear: 100~s$^{-1}$). This flow curve does not display, however, any yield stress, in agreement with the macroscopic observations of the dispersion in its vial. Similar flow curves were reported for dispersions below but close to a glass transition~\cite{Fuchs2005}.
The red dots in Fig.~\ref{fig:RheologySI.pdf}b are measurements of stress at several imposed low shear rates after the ramp.  We extract from these measurements a viscosity at low shear rates $\eta \simeq 6 \pm 2$~Pa.s. Note that we performed several experiments on this concentrated sample, all displaying rather similar behaviors, but with a high variability of the estimated low shear viscosity. We believe that these variations could be explained by unavoidable small evaporation during the rheological measurements, combined with the very high susceptibility of $\eta$ with $\varphi$ close to the glass transition. All of these viscosity measurements as a function of the volume fraction enables us to plot the curve provided in Fig.~\ref{fig:FlowCurve}a.

\balance

\renewcommand\refname{References}


\providecommand*{\mcitethebibliography}{\thebibliography}
\csname @ifundefined\endcsname{endmcitethebibliography}
{\let\endmcitethebibliography\endthebibliography}{}

\end{document}